\documentclass[sigconf]{acmart}
\AtBeginDocument{%
  \providecommand\BibTeX{{%
    \normalfont B\kern-0.5em{\scshape i\kern-0.25em b}\kern-0.8em\TeX}}}

\setcopyright{acmcopyright}
\copyrightyear{2018}
\acmYear{2018}
\acmDOI{XXXXXXX.XXXXXXX}

\acmConference[Conference acronym 'XX]{Make sure to enter the correct
  conference title from your rights confirmation emai}{June 03--05,
  2018}{Woodstock, NY}
\acmPrice{15.00}
\acmISBN{978-1-4503-XXXX-X/18/06}

\begin{document}

\title{Linking the Dynamic PicoProbe Analytical Electron-Optical Beam Line / Microscope to Supercomputers}

\author{
Alexander Brace$^{1,2}$, 
Rafael Vescovi$^{1}$,
Ryan Chard$^{1}$,
Nickolaus D. Saint$^{2}$,
Arvind Ramanathan$^{1,2*}$,
Nestor J. Zaluzec$^{3,4}$,
Ian Foster$^{1,2*}$
}

\affiliation{
$^{1}$Data Science and Learning Division, Argonne National Laboratory, Lemont, IL, USA. 
$^{2}$Computer Science Department, University of Chicago, Chicago, IL, USA.
$^{3}$Photon Sciences Division, Argonne National Laboratory, Lemont, IL, USA.
$^{4}$Physical Sciences Division, University of Chicago, Chicago, IL, USA. \\
Contact authors: \{ramanathana, foster\}@anl.gov
\country{}
}
% \author{Alexander Brace}
% \email{abrace@uchicago.edu}
% \orcid{0000-0001-9873-9177}
% \affiliation{
%   \institution{University of Chicago and Argonne National Laboratory}

% \author{Rafael Vescovi}
% \email{ravescovi@anl.gov}
% \orcid{0000-0002-0652-1128}
% \affiliation{
%   \institution{Argonne National Laboratory}

% \author{Ryan Chard}
% \email{rchard@anl.gov}
% \orcid{0000-0002-6781-7432}
% \affiliation{
%   \institution{Argonne National Laboratory}

% \author{Nickolaus D. Saint}
% \email{nickolaus@uchicago.edu}
% \orcid{0000-0002-1484-8585}
% \affiliation{
%   \institution{University of Chicago}

% \author{Arvind Ramanathan}
% \email{ramanathana@anl.gov}
% \orcid{0000-0002-1622-5488}
% \affiliation{
%   \institution{Argonne National Laboratory}

% \author{Nestor J. Zaluzec}
% \email{zaluzec@anl.gov}
% \orcid{0000-0002-4857-3487}
% \affiliation{
%   \institution{University of Chicago and Argonne National Laboratory}

% \author{Ian T. Foster}
% \email{foster@anl.gov}
% \orcid{0000-0003-2129-5269}
% \affiliation{
%   \institution{University of Chicago and Argonne National Laboratory}
\renewcommand{\shortauthors}{Brace, et al.}

% TODO: Add quantitative metric
\begin{abstract}
The Dynamic PicoProbe at Argonne National Laboratory is undergoing upgrades that will enable it to produce up to 100s of GB of data per day. While this data is highly important for both fundamental science and industrial applications, there is currently limited on-site infrastructure to handle these high-volume data streams. We address this problem by providing a software architecture capable of supporting large-scale data transfers to the neighboring supercomputers at the Argonne Leadership Computing Facility. To prepare for future scientific workflows, we implement two instructive use cases for hyperspectral and spatiotemporal datasets, which include: (i) off-site data transfer, (ii) machine learning/artificial intelligence and traditional data analysis approaches, and (iii) automatic metadata extraction and cataloging of experimental results. This infrastructure supports expected workloads and also provides domain scientists the ability to reinterrogate data from past experiments to yield additional scientific value and derive new insights.
\end{abstract}




%% TODO: Update these?
\received{20 February 2007}
\received[revised]{12 March 2009}
\received[accepted]{5 June 2009}

\maketitle

\section{Introduction}
Experimental facilities around the world rely on computational infrastructure to support scientific discovery. Such infrastructure is present at all levels of the ``experimental stack'', including: precise control of instrumentation for atomic-scale measurement, capturing and analyzing big data in real-time, and publishing such data for the broader research community to access. An increasingly important design pattern within this paradigm, ``closing the loop,'' seeks to tighten the gap between experiment and computation, thereby increasing the efficiency of the research process and accelerating the rate of discovery. Machine learning and artificial intelligence (ML/AI) play a lead role in this pattern as the underlying computational agents that, in many scenarios, are able to optimize a set of experimental measurements toward an objective. To achieve this vision, there is a need for robust, open-source, modular software components to realize end-to-end experimental workflows and open the door for computationally mediated science.

\begin{figure*}[t!]
  \centering
  \includegraphics[width=\textwidth]{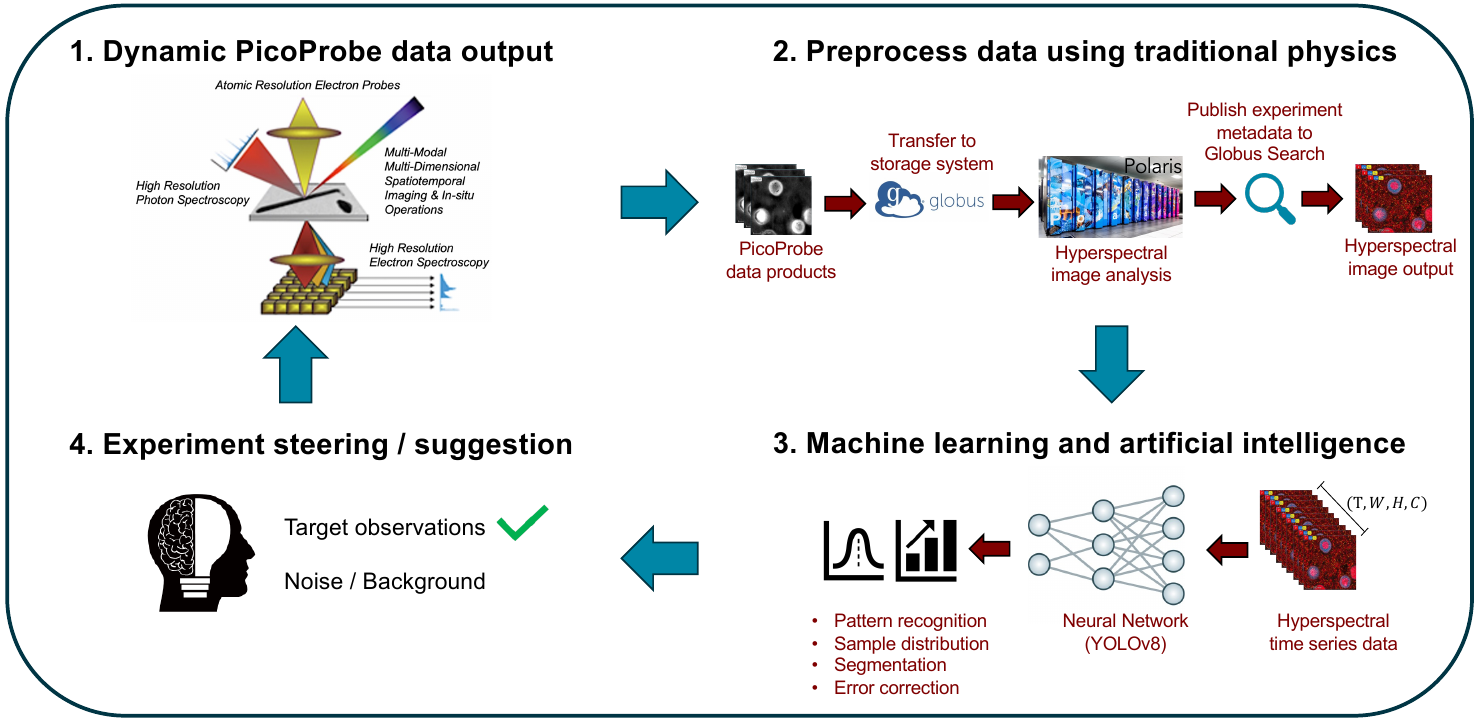}
  \caption{A high-level vision to support computationally mediated science at the Dynamic PicoProbe. (1) High-dimensional data is generated by the Dynamic PicoProbe. (2) Data is transferred from on-site computers to more powerful computing clusters (e.g., ALCF's Polaris) to perform hyperspectral analysis using traditional approaches and metadata tracking to enable scientists to later reinterrogate data. (3) The hyperspectral data is used as input to ML/AI approaches to: (i) discover interesting patterns in the data, (ii) characterize the data distribution, (iii) segment and detect features in the data to assist in calibrating measurement, (iv) perform error correction by alerting the Dynamic PicoProbe operator to calibration problems. (4) The data analysis and metadata are synthesized into an actionable summary to assist domain scientists in performing the measurement(s) (i.e., experiment) of interest. In this work, we present steps (2) and (3) as different flow use cases, rather than a single unified flow.
}
  \Description{}
  \label{fig:overview}
\end{figure*}

In this work, we describe our approach to developing such infrastructure for the Dynamic PicoProbe Analytical Electron-optical Beam Line / Microscope (Sec.~\ref{Methods:Dyanamic-PicoProbe}) at Argonne National Laboratory (ANL). The Dynamic PicoProbe is undergoing a set of upgrades, and upon completion is expected to produce 100s of GB of data per day during steady-state operation. In the long term, future state-of-the-art detectors (which will further extend scientific capabilities) will generate up to 65 GB of data per second ($\approx$200 TB/hour). To prepare for these intensive data streams, we employed Globus automation services~\cite{chard2023globus,vescovi2022linking} to develop a pair of data flows for transferring experimental data (hyperspectral and spatiotemporal images) from the Dynamic PicoProbe site to the Polaris supercomputer at the Argonne Leadership Computing Facility (ALCF) for analysis and cataloging, as illustrated in Fig.~\ref{fig:overview}. In addition to transferring data into a permanent store at ALCF, we provide a simple access portal for researchers to view experimental analyses to help guide the next set of experimental measurements and easily share their findings. We leveraged the Globus Search service and the Django Globus Portal Framework (DGPF)~\cite{saint2023active} to make data Findable, Accessible, Interoperable, and Reusable (FAIR)~\cite{wilkinson2016fair}.

This work presents computational infrastructure to facilitate automated data transfer from the Dynamic PicoProbe to the Argonne Leadership Computing Facility (ALCF) in order to:
\begin{itemize}
    \item Provide a crucial data storage solution for experimental data that would quickly overwhelm on-site computing resources.
    \item Use high performance computing to (i) analyze hyperspectral data, and (ii) leverage ML/AI on spatiotemporal data streams to extract/summarize scientifically meaningful information from high-volume and high-velocity data streams.
    \item Produce a FAIR search index and user portal to catalog experiment metadata and data products to support scientific campaigns over extended durations and with multiple users.
\end{itemize}

Our code, ML/AI model, and datasets are open-source and freely available on GitHub\footnote{https://github.com/ramanathanlab/PicoProbeDataFlow}.

\section{Methods}

\subsection{Dynamic PicoProbe}
\label{Methods:Dyanamic-PicoProbe}
Upgrades to the Dynamic PicoProbe~\cite{zaluzec2021first} at ANL will enable multi-modal, multi-dimensional, and in-situ characterization of dynamic events at interfaces in environmental media. Its capabilities will include temporally resolved (< 3 ms)~\cite{scheres2012bayesian} and sub-atomic ($\sim$ 50 pm)~\cite{venkatakrishnan2014model} hyperspectral imaging during in-situ/operando operations in high-vacuum, cryogenic, liquid, and/or gaseous environments of macromolecular/ionic species of beam-sensitive and soft/hard matter systems. The Dynamic PicoProbe features: (i) a 30-300 kV monochromated, aberration-corrected 50 pm electron probe; (ii) sub-atomic imaging capabilities; (iii) high energy resolution electron spectroscopy (< 30 meV); and (iv) the $X_{PAD}$, the world's highest collection efficiency hyperspectral x-ray detector array ($\sim$4.5 sR).

The Dynamic PicoProbe is controlled by a host computer running Windows 10. Commands are issued through a GUI and haptic control panel. Data from the instrument are relayed back to the host computer for interactive control and visualization from four Linux and two Windows 10 systems that control data acquisition. Windows 10 and macOS user workstations facilitate data processing, external data transfers, and data backups of user-curated images and data products (e.g., hyperspectral images). Currently, user machines are equipped with a 1 Gbps switch that handles external data transfers. Upgrades are underway to route data directly from the data acquisition system to the Argonne National Laboratory backbone, which runs at up to 200 Gbps on-site.

\subsection{Data Flow Infrastructure}
\label{Methods:Data-flow-Infrastructure}
This section describes the computational infrastructure that we use to transfer, analyze, and publish experimental data in near real-time. Each data flow in this work comprises three distinct processing steps: (i) \textit{Data Transfer} with Globus, (ii) \textit{Data Analysis}, whereby data products are analyzed, plots are produced, and experiment metadata are extracted, and (iii) \textit{Data Publication}, in which the generated plots and experiment metadata are published to a Globus Search index. We use the term ``flow'' as a shorthand to describe a data flow in which multiple stages of computation run serially across heterogeneous resources and locations. We implement our flows in Python by using the Globus Architecture for Data-Intensive Experimental Research (Gladier) software package~\cite{vescovi2022linking}.

\subsubsection{Data Transfer}
\label{Methods:DataTransfer}
Before a data flow can begin, new data produced from an experiment must be automatically recognized and used to invoke the flow. While the scientific use cases highlighted here (Sec.~\ref{Results:Hyperspectral-Imaging-Data-flow}, ~\ref{Results:Spatiotemporal-Imaging-Data-flow}) focus on user-curated data files, we can apply the same design principles to process data files written directly by the experimental instrument software. To support automatic data transfers, we developed a cross-compatible Python application for Windows 10, macOS, and Linux that uses the watchdog package~\cite{watchdog} to start a new flow when files are created on the user machine (Sec.~\ref{Methods:Dyanamic-PicoProbe}). Our application is very lightweight as the task logic, orchestration, and fault tolerance are managed by Gladier/Globus automation services. This software stack allows scaling the number of concurrent flows (as supported by the available networking infrastructure) to keep pace with the data-velocity. We also provide an automatic checkpointing mechanism to avoid undesired flow repeats in cases where a user needs to resume experimentation after interruption, e.g., if the user computer needs to be rebooted or the user resumes a set of experiments on a subsequent day.

When a new file is detected, the Python application starts a Globus flow. Upon flow start, files are transferred from the user computer to ALCF's Eagle storage system, an O(100PB) Lustre file system. In our example use cases, the files are written in the Electron Microscopy Dataset (EMD) format, a subset of the Hierarchical Data Format version 5 (HDF5) format that efficiently stores high dimensional microscopy data (including hyperspectral images and spatiotemporal images) in a standardized, efficient, binary format. Provisions are also incorporated to use other cross-platform formats such as the proposed ISO standard HMSA format~\cite{hmsa_spec}, as well as parallel formats used in the scientific community. The 
data is moved by using the Globus Transfer service, a cloud-hosted solution for copying data rapidly and reliably between Globus Connect endpoints. Transfer leverages the OAuth-based Globus Auth to identify and authenticate users to ensure data is moved securely.

\subsubsection{Data Analysis}
\label{Methods:DataAnalysis}
Once the EMD files arrive on the Eagle file system, two computational steps are performed: %in the current implementation: 
(i) image processing, and (ii) experiment metadata extraction. 

For image processing, we employ Globus Compute~\cite{chard2020funcx}, a federated function-as-a-service platform for secure and reliable remote computation, to request a compute node on ALCF's Polaris supercomputer and thus avoid overwhelming the login nodes. The Globus Compute model employs user-deployed endpoint agents on remote resources to perform tasks. A user may submit Python functions for execution by specifying the function body, arguments, and the endpoint on which the code is to be executed. The Globus Compute service securely routes the task to the endpoint, where it may either provision batch resources or perform the task locally before results are returned to the user via the Compute service. In our case, the endpoint is configured to acquire compute nodes on the Polaris supercomputer by using the PBS scheduler.

Next, the EMD file is parsed to extract experiment metadata by using the HyperSpy Python package~\cite{francisco_de_la_pena_2022_7263263}. The metadata includes sample collection date and time; acquisition instrument (i.e., microscope) details, such as stage and detector positions, beam energy, and magnification; and other information, such as software versioning. To increase the end-to-end efficiency of the flow, we combine metadata extraction and the image processing steps into a single Globus Compute function which avoids reading the EMD file twice and minimizes flow orchestration overhead.

\subsubsection{Data Publication}
\label{Methods:DataPublication}
Data publication is achieved by creating and registering the data and associated metadata (defined by using an extensible schema based on DataCite~\cite{brase2009datacite}) with a Globus Search index. Globus Search is a cloud-hosted service that builds on ElasticSearch to enable users to create, populate, and manage indices of searchable metadata. Search provides a fully-featured free-text search model along with fine-grained security and access control to facilitate visibility-filtered query results that restrict data discoverability to authorized users. The publication process is a light-weight action that transmits the JSON metadata, extracted during the \textit{Data Analysis} step (Sec.~\ref{Methods:DataAnalysis}), to the Search service, and can be performed on a Polaris login node. 

Metadata and results are then visualized by using a Django Globus Portal Framework (DGPF) data portal. DGPF combines the Django web framework with Globus to create a customizable data portal based on the Modern Research Data Portal, a design pattern for providing secure, scalable, and high performance access to research data. DGPF portals can be used to dynamically display records in a Globus Search index while leveraging the trusted authentication systems to render data and results hosted on remote Globus Connect endpoints. DGPF catalogs have been used to index millions of datasets consisting of many terabytes of data. In this work, we enable researchers to search their experimental data and results by the time and date of the associated experiment.

\section{Results}
We present results for two scientific use cases, (i) \textit{hyperspectral imaging}, and (ii) \textit{spatiotemporal imaging}, showcasing the generated results from each application via a user portal display. We also provide a brief vignette that illustrates the current performance hurdles and suggests areas for improvement.
 
\subsection{Hyperspectral Imaging Data Flow}
\label{Results:Hyperspectral-Imaging-Data-flow}
\begin{figure*}[ht]
  \centering
  \includegraphics[width=\linewidth]{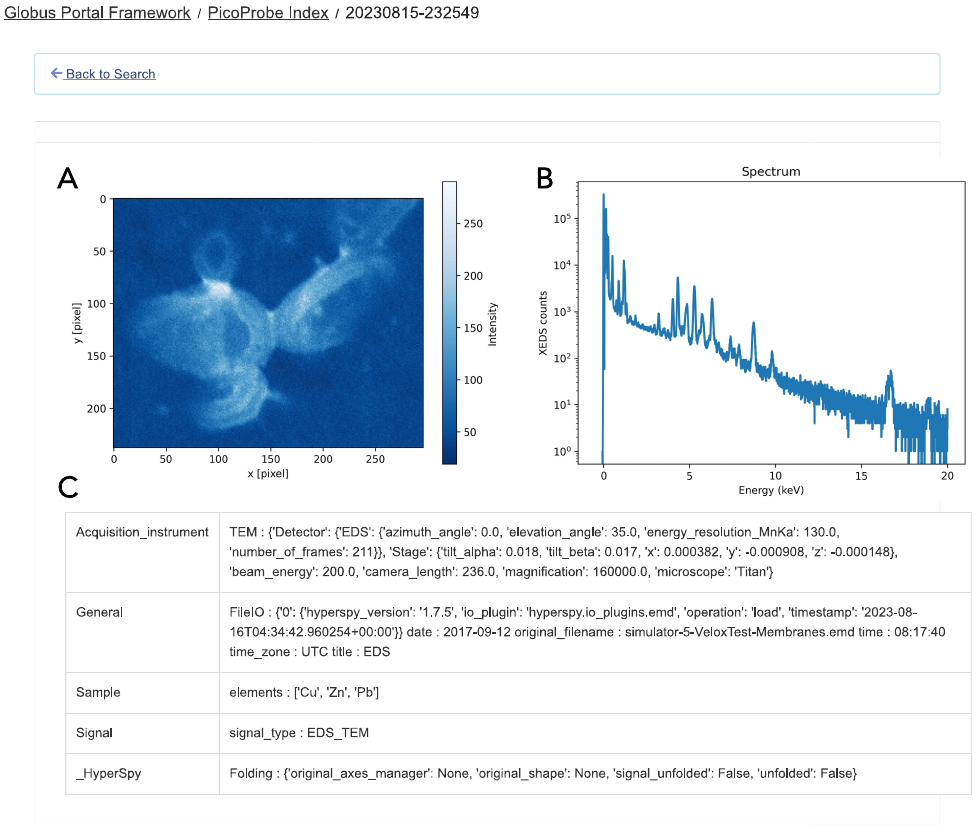}
  \caption{The DGPF interactive user portal allows researchers to quickly access experimental results and metadata backed by Globus Search. Here we show (A) a hyperspectral image of a polyamide organic film treated to capture heavy metals from water~\cite{song2019unraveling}, (B) its corresponding spectrum, and (C) metadata indicating the microscope settings used to collect the data as well as the atomic composition of the sample.}
  \Description{}
  \label{fig:hyperspectral}
\end{figure*}

The hyperspectral image is processed as a 3-dimensional tensor containing pixel-width, pixel-height, and hyperspectral data. The hyperspectral data comprises spectroscopic information about both the atomic elemental composition present in the sample at a given pixel location, as well as the electron scattering/image data. We generate a plot for users by taking a sum along the spectroscopy dimension to compute the intensity of the sample at each pixel, depicted in Fig.~\ref{fig:hyperspectral}.A. We also generate a plot of the entire sample's spectrum by summing the image over each of the pixel dimensions, as shown in Fig.~\ref{fig:hyperspectral}.B. This spectrum conveys information about the aggregate atomic composition present throughout the sample. These plots are automatically rendered in the DGPF data portal along with the extracted experimental metadata: see Fig.~\ref{fig:hyperspectral}.C.

\subsection{Spatiotemporal Imaging Data Flow}
\label{Results:Spatiotemporal-Imaging-Data-flow}
As a hierarchical file format, EMD files can store many forms of microscopy data. In this use case, each EMD file stores a multi-dimensional spatiotemporal image tensor, where the first axis stores the time dimension and the inner axes store the pixel-width and pixel-height of the image signal. Here, we investigate a spatiotemporal image with 600 frames showing the motion of gold nanoparticles on a carbon background. Notably, an additional hyperspectral dimension (Sec.~\ref{Results:Hyperspectral-Imaging-Data-flow}) could be added which would result in a 4-dimensional tensor, vastly increasing the data volume of each file---we leave this use case to future work.

\begin{figure}[ht]
  \centering
  \includegraphics[width=0.9\columnwidth]{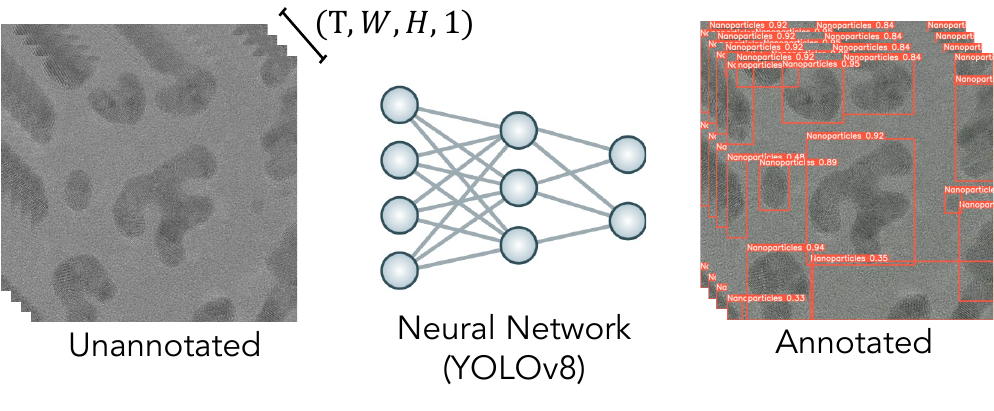}
  \caption{Spatiotemporal data is preprocessed so that each time step is input to a fine-tuned YOLOv8 model to track locations of gold nanoparticles. The bounding boxes (orange) have a confidence score and can be used to count the number of nanoparticles likely to be in a sample, helping to characterize changes in the sample as a function of time.}
  \Description{}
  \label{fig:spatiotemporal}
\end{figure}

In order to prepare for such data streams, we demonstrate ML/AI approaches to automatically track scientifically meaningful information about sample contents. Specifically, we train a YOLOv8~\cite{yolov8_ultralytics} model to detect and track gold nanoparticles/nanostructures as they move. Before training a YOLOv8 model, hand-labeled bounding boxes must be drawn around the prediction targets (in this case, a single label type representing the gold nanoparticles). As the dataset presented in this work has 600 time steps, we select every 50th step for hand-labeling with Roboflow~\cite{dwyer2022roboflow}, yielding a total of nine training, three validation, and one testing 640$\times$640 image(s). We further augmented the training set by using horizontal and vertical flips, as well as random cropping up to 20\% maximum zoom. We then fine-tuned the YOLOv8s model (11.2M parameters) in Google Colab for 100 epochs with stochastic gradient descent, with a batch size of 16 and a learning rate of 0.01, on a Tesla T4 GPU. Our model achieves a mean Average Precision with an Intersection over Union (IoU) range of 50--95\% (mAP50-95) of 0.791 on the training set and 0.801 on the validation set.

We employ the fine-tuned YOLOv8 model for efficient inference within the spatiotemporal imaging data flow by first converting incoming EMD files to MP4 video format, followed by calling the inference routine in a subprocess. We performed inference on a Polaris compute node with an NVIDIA A100 GPU and output an annotated MP4 file containing the predicted gold nanoparticle bounding boxes, as illustrated in Fig.~\ref{fig:spatiotemporal}.

\subsection{Performance Evaluation}
To provide a controlled environment for testing our data flow infrastructure, we employ an application that periodically copies a file into the transfer directory of the Dynamic PicoProbe user computer to simulate data generation during an actual experiment. We configure the experiments based on the approximate time it takes each transfer to complete. Thus, over the course of an hour, we automatically start a new flow every 30 and 120 seconds for the hyperspectral and spatiotemporal use cases, respectively. The Globus services allow parallel flow execution that enables us to start new flows even when previous ones are still running. We summarize the performance metrics for our use cases in Table~\ref{tab:performance}. Note that the file size in the hyperspectral use case (91 MB) is much smaller than the spatiotemporal counterpart (1200 MB), which leads to many more hyperspectral flow runs completing within the allotted hour. The maximum runtimes are associated with the first flows, as they have to request a compute node on Polaris and cache the Python libraries required for analysis. Subsequent flows are able to reuse nodes already provisioned to the previous flows.

\begin{table}
\centering
\caption{Performance measured during independent 1-hour long experiments in which files were transferred from the Dynamic PicoProbe user computer to the Eagle filesystem.}
\begin{tabular}{@{}lcc@{}}
\toprule
\textbf{Metric} & \textbf{Hyperspectral} & \textbf{Spatiotemporal} \\
\midrule
Start period (s) & 30 & 120 \\
Transfer volume (MB) & 91 & 1200 \\
Total data transfer (GB) & 6.42 & 21.72 \\
Min flow runtime (s) & 29 & 195 \\
Mean flow runtime (s) & 47 & 224 \\
Max flow runtime (s) & 181 & 274 \\
Median overhead (s) & 19.5 & 45.2 \\
Median overhead (\%) & 49.2 & 21.1 \\
Total flow runs & 72 & 18 \\
\bottomrule
\label{tab:performance}
\end{tabular}
\end{table}

\begin{figure}[hb]
  \centering
  \includegraphics[width=\linewidth]{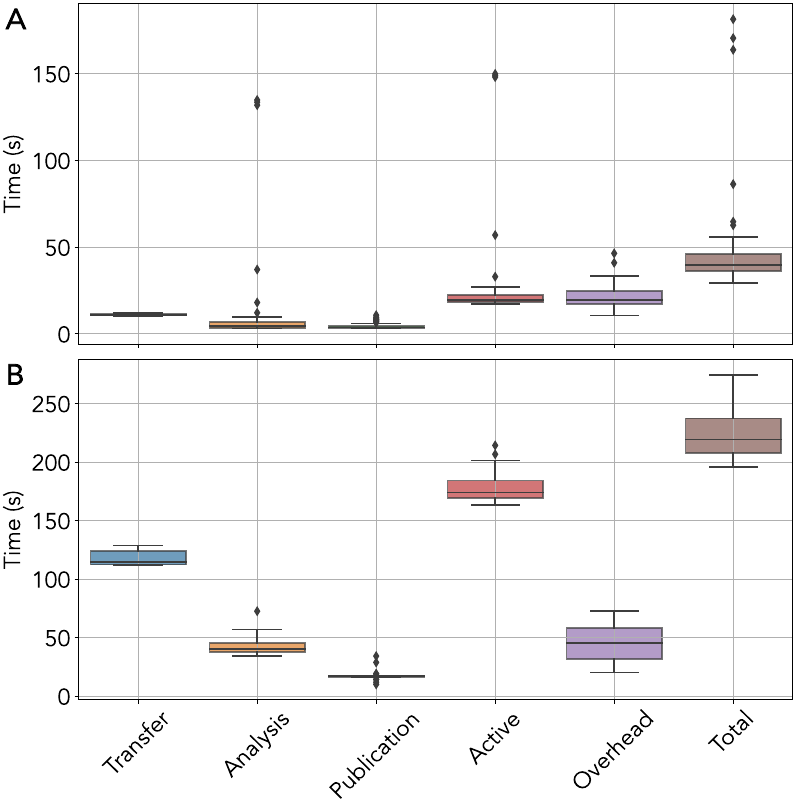}
  \caption{The itemized runtime statistics (in seconds) for the hyperspectral flow (A) and the spatiotemporal flow (B) measured over independent 1-hour long experiments. We use ``Active'' to denote the time spent actively processing either the Transfer, Analysis, or Publication steps. The overhead measures the remainder of the total flow runtime not spent actively processing the steps.}
  \Description{}
  \label{fig:flow-runtime}
\end{figure}

In addition to aggregate flow statistics, we characterize the performance of the individual flow component steps (Sec.~\ref{Methods:Data-flow-Infrastructure}) to profile and understand performance bottlenecks. We illustrate the runtime statistics (in seconds) of the hyperspectral flow in Fig.~\ref{fig:flow-runtime}.A, and spatiotemporal flow in Fig.~\ref{fig:flow-runtime}.B, as well as the time spent actively processing tasks versus the overhead. The flow orchestration overhead is significant at 49.2\% of the total median runtime for the hyperspectral flow and 21.1\% for the spatiotemporal flow. This observation is attributed to an exponential polling backoff policy that starts at 1 second and doubles up to 10 minutes, which we are working to improve. Outside of overhead, the file transfer time, which dominates active flow runtime, is primarily a function of the size of the transferred files and demonstrates the need to update on-site data transfer capabilities (currently facilitated by a 1~Gbps switch (Sec.~\ref{Methods:Dyanamic-PicoProbe})) to support future detectors, which will produce up to 65 GB per second. The spatiotemporal compute phase can also be optimized since the majority of time is spent on converting raw EMD files to MP4 format, which involves a slow data type casting operation from fp64 to uint8. More efficient integration with the YOLOv8 algorithm would lead to a substantial improvement in time-to-solution for spatiotemporal data stream analysis.

\section{Related Work}
\label{RelatedWork}
We distinguish in this work between \textit{computationally mediated science} and \textit{automated science}. In the former, automation and intelligent agents augment the abilities of a human scientist by performing routine tasks, automatic calibrations, and sharing in a collaborative synergy to accelerate discovery~\cite{zaluzec2003computationally}. On the other hand, automated science seeks to create self-driving laboratories that integrate a variety of instruments under the direction of an ML/AI agent, which automatically steers an experimental campaign towards discoveries using minimal human intervention~\cite{vescovi2023towards, king2009automation}. These directions share a common need for modular software infrastructure~\cite{balouek2019towards, salim2019balsam, ananthakrishnan2020open, vescovi2022linking, zhang202120} to link experimental facilities to high performance computing resources~\cite{wang2001high, veseli2018aps, basu2019automated, liu2021bridging}.

ML/AI has been successfully applied to a variety of electron microscopy tasks including: detecting microstructures~\cite{zhong2022machine}, image registration~\cite{luo2021real}, pixel-level nanoparticle segmentation~\cite{larsen2022large}, frame-level defect tracking and detection~\cite{shen2021deep}, and many other applications such as particle picking~\cite{wagner2019sphire}, automated labeling~\cite{weber2018automated}, denoising~\cite{bepler2020topaz}, and super-resolution reconstruction~\cite{suveer2019super}. 
Treder et al.~\cite{treder2022applications} provide a comprehensive review of deep learning in electron microscopy. Our work extends these efforts by establishing the infrastructure needed to bridge microscopes with HPC infrastructure for online integration of computationally expensive ML/AI and traditional analysis techniques.

\section{Conclusion}
We have presented software infrastructure that links the Dynamic PicoProbe to supercomputing resources to (1) provide petabyte-scale data storage, (2) enable near real-time data analysis by using ML/AI techniques, and (3) build interactive FAIR data portals for researchers to view results and inform future experiments. We leverage Globus automation services to implement two prototypical science use cases (hyperspectral imaging and spatiotemporal imaging), providing configurable software that decouples data analysis steps from the limitations of on-site computers. We found transfer times between on-site data staging systems and the supercomputing facility to be the overall bottleneck in our data flows. We expect this issue to compound as data tensor dimensions are introduced to measure additional physical parameters such as temperature and pressure. As such, active directions for future research include: (1) on-site hardware upgrades, (2) data compression algorithms, and (3) optimization of cross-site transfer settings.

\begin{acks}
The Analytical Picoprobe at Argonne National Laboratory (ANL) was developed as part of CRADA \#01300710 between ANL and ThermoFisher Scientific Instruments.  This work is supported by the U. S. Department of Energy, Office of Science, Office of Basic Energy Sciences, under Contract No. DE-AC02-06CH11357, as well as the National Science Foundation Major Research Instrumentation (MRI) Program (NSF DMR-2117896) at the University of Chicago. This work was supported in part by NSF grants OAC-1835890 and OAC-2004894. We are grateful to the Globus team for their support. We thank Carla M. Mann for helping to proofread this manuscript.
\end{acks}

\bibliographystyle{ACM-Reference-Format}
\bibliography{references}

% \appendix

\end{document}